\documentclass[12pt,a4paper]{article}
\usepackage[height=22.5cm,width=16.5cm]{geometry}
\usepackage{amsmath,amssymb}
\usepackage[mathscr]{eucal}
\usepackage{hyperref}
\usepackage{cite}
\usepackage{bm}
\usepackage{braket}
\numberwithin{equation}{section}

\def\cM{\mathcal{M}}
\def\cS{\mathcal{S}}
\def\cN{\mathcal{N}}
\def\PD{\mathrm{PD}}
\def\red{\text{red}}
\def\dd{\mathrm{d}}
\begin{document}
\begin{titlepage}

\begin{flushright}
IPMU 13-0067 \\
UT-13-12
\end{flushright}
\vskip 2cm

\begin{center}
{\Large \bfseries
Notes on reductions of superstring theory\\[.5em]
 to bosonic string theory
}

\vskip 1.2cm

Kantaro Ohmori$^\flat$
and Yuji Tachikawa$^{\flat,\sharp}$

\bigskip
\bigskip

\begin{tabular}{ll}
$^\flat$  & Department of Physics, Faculty of Science, \\
& University of Tokyo,  Bunkyo-ku, Tokyo 133-0022, Japan,\\
$^\sharp$  & Kavli Institute for the Physics and Mathematics of the Universe, \\
& University of Tokyo,  Kashiwa, Chiba 277-8583, Japan
\end{tabular}

\vskip 1.5cm

\textbf{Abstract}
\end{center}

\medskip
\noindent

It is in general  very subtle  to  integrate over the odd  moduli of super Riemann surfaces in perturbative superstring computations.
We study how these subtleties go away in favorable cases, including the  embedding of $\cN=0$ string to $\cN=1$ string by Berkovits and Vafa, and the relation of the graviphoton amplitude and the topological string amplitude by Antoniadis, Gava, Narain and Taylor and Bershadsky, Cecotti, Ooguri and Vafa.
The Poincar\'e dual of the moduli space of  Riemann surfaces in the moduli space of super Riemann surfaces plays an important role.
 
\bigskip
\vfill
\end{titlepage}

\section{Introduction and Summary}
Perturbative superstring  amplitudes are given by an integral over the moduli space $\cM$ of super Riemann surfaces\cite{Friedan:1985ge,Verlinde:1987sd,Atick:1987wt,D'Hoker:2002gw}.  
As clarified in a series of papers last year \cite{Witten:2012bg,Witten:2012ga,Witten:2012bh},
general properties of superstring amplitudes can be formulated and be understood 
directly in terms of $\cM$.
That said, practical computations were often done by first reducing the integral on $\cM$ to an integral
on  the moduli space $\cM_\red$ of Riemann surfaces%
\footnote{In this article we use $X_\red$ to denote the bosonic part of the supermanifold $X$, following \cite{Witten:2012bg}.}, 
using a projection 
\begin{equation}
p: \cM\to \cM_\red. \label{projection}
\end{equation} 
A main  source of the difficulties is that $p$ cannot in general  be globally holomorphic 
when the genus is sufficiently high \cite{Donagi:2013dua}. 
This fact introduces various subtle global issues when one tries to first integrate over the odd moduli.
This problem was often called the ambiguity of the integrand in the literature, see e.g.~\cite{Atick:1987wt}. 
The viewpoint emphasizing the supermoduli space clarifies this ambiguity, and also provides a way to deal with it in a consistent manner, as reviewed below.

Yet there are favorable cases where superstring amplitudes can be reduced to bosonic string amplitudes by showing a relation schematically of the form \begin{equation}
\int_{\cM} F = \int_{\cM_\red} F'. \label{aho}
\end{equation} Two examples are \begin{itemize}
\item the embedding of an arbitrary bosonic string to $\cN=1$ superstring by Berkovits and Vafa\cite{Berkovits:1993xq}, and
\item the reduction of the graviphoton amplitudes in Calabi-Yau compactifications to the topological string amplitudes by Antoniadis, Gava, Narain and Taylor \cite{Antoniadis:1993ze} and by Bershadsky, Cecotti, Ooguri and Vafa \cite{Bershadsky:1993cx}. Note that topological strings are bosonic strings as far as the integration over the moduli space is concerned.
\end{itemize}

The aim of this short note is to show that the so-called ambiguity of the integrand does not affect these equivalences, because only the holomorphic inclusion  \begin{equation}
\iota: \cM_\red \to \cM\label{inclusion}
\end{equation} is used in the analysis. Namely, we will find below that the integrand $F$ on the superstring side automatically has the form \begin{equation}
F=F'\wedge \PD[\cM_\red]\label{PD}
\end{equation} where $\PD[\cM_\red]$ denotes the Poincar\'e dual of the bosonic moduli space in the super moduli space.  This structure is a consequence of a simple argument involving charge conservation on the world sheet.

In the formalism using the picture changing operators (PCOs),\cite{Friedan:1985ge,Verlinde:1987sd,Atick:1987wt}, our observation can be summarized as follows.
In a patch of the supermoduli space, we can integrate out the odd moduli, producing the insertion of PCOs. 
Changing the positions of the PCOs in general produces an exact form on the bosonic moduli space. 
This leads to various ambiguities in the integral when we glue together the contributions from various patches to obtain the total amplitude. 
The observation of \cite{Witten:2012bg,Witten:2012ga,Witten:2012bh} is that this ambiguity is in general unavoidable if we insist on trying to reduce the computation on the integral over the bosonic moduli space, and that the computation free of ambiguity is possible if we consider the integral over the supermoduli space  as it is. 
Therefore, when computing general amplitudes, it is not a very good idea to start from the integration over the bosonic moduli space of correlation functions with PCOs and then to worry the possible ambiguities.

What we are going to show in this note is that
the two favorable cases mentioned above, and in general the condition \eqref{PD}, correspond to the cases where those exact forms are zero. In the language using PCOS, we find that the correlation functions are completely independent of their positions.
In the case of embedding of the $\mathcal{N}=0$ string to $\mathcal{N}=1$ string, this independence was already mentioned in the original paper\cite{Berkovits:1993xq}. The independence on the positions of PCOs in the graviphoton amplitude is a new observation.

Differently put, the two cases mentioned above and in general the case covered by \eqref{PD} are very special in that the integral over the supermoduli space naturally reduces to an integral over the bosonic moduli space.  Therefore we use a more general framework of \cite{Witten:2012bg,Witten:2012ga,Witten:2012bh} in this note, explaining the correspondence with the more traditional framework of \cite{Friedan:1985ge} in appropriate places. For example, using PCOs corresponds to taking the worldsheet gravitino to be supported by delta functions.  Our results applies to more  general gauge choices of the  worldsheet gravitino; the only condition is that the gravitino is zero where the physical vertex operators sit. 

The rest of the article is organized as follows. In Sec.~\ref{review} we review the basics of the superstring perturbation theory as an integration over the moduli space of  super Riemann surfaces. In particular we recall where the difficulties from the integration of odd moduli arise. In Sec.~\ref{varpi} we explain when these difficulties do not arise, thanks to the appearance of the Poincar\'e dual of the reduced space in the superspace.  In Sec.~\ref{BV} we apply this analysis to the embedding of the $\cN=0$ string to $\cN=1$ string by Berkovits and Vafa, and in Sec~\ref{topological} the same analysis is similarly applied to the reduction of the graviphoton amplitudes to the topological string amplitude. 
The relation of the approach used in this note and the more traditional formalism using the PCOs are reviewed and explained in Appendix~\ref{fail}.

We note that the contents of Sec.~\ref{varpi} and of Sec.~\ref{BV} and much more were independently reached by Witten \cite{WittenEmbedding}.

\section{A short review on integrals over supermoduli}\label{review}

In this section we collect the basics of supermanifolds and super Riemann surfaces to formulate superstring perturbation theory. Those readers who are familiar with the contents of \cite{Witten:2012bg,Witten:2012ga,Witten:2012bh} can directly skip to the next section.

\subsection{Supermanifolds and super Riemann surfaces}
A supermanifold of dimension $p|q$, where $p$ is the bosonic dimension and   $q$ is the fermionic dimension,
is constructed by pasting subsets of superspace  $\mathbb{R}^{p|q}$.
Two patches represented by coordinates $(x^i|\theta^\mu)$ and $(x^{i\prime}|\theta^{\mu\prime})$ are glued together with gluing functions:
\begin{equation}
	x^{i^\prime}=f^i(x|\theta),\quad
	\theta^{\mu\prime}=\psi^\mu(x|\theta)
	\label{gluing}
\end{equation}
where $f$ is even and $\psi$ is odd. 
If $f$ and $\psi$ are both real, then the resulting supermanifold $M$ is called real.
If the constraint is only $f(x|\theta=0)$ is real, then $M$ is called cs manifold.
If we replace $\mathbb{R}^{p|q}$ with $\mathbb{C}^{p|q}$ and require $f$ and $\psi$ to be holomorphic , we get complex supermanifold.

We can construct an ordinary manifold $M_\text{red}$, called the reduced manifold, form supermanifold $M$ by ignoring all fermionic things.  Gluing functions of $M_\text{red}$ is simply $f^i(x|\theta=0)$.
Reduced manifolds of real or cs supermanifolds are real manifolds, 
and reduced manifolds of complex supermanifolds are complex manifolds.

If we can take coordinates such that $f$ of (\ref{gluing}) depend only on bosonic coordinates, the manifold called projected.
All smooth supermanifolds are smoothly projected %\cite{Batchelor:1979}
 but in general complex supermanifolds  are not holomorphically projected.
This prevents us from naively dealing with odd moduli of superstring worldsheets.

We have to integrate superfunctions over supermanifold to set up superstring perturbation theory.
We  define an integral top form as $\Omega=f(x|\theta)\prod_i\delta(\dd x^i)\prod_\mu\delta(\dd \theta^\mu)$ where $\dd x^i=\delta(\dd x^i)$ for bosonic variables $x^i$.
The integration of an integral top form $\Omega$ over a patch $U_\alpha$ is defined as:
\begin{align}
	\int_{U_\alpha}\left.\Omega\right|_{U_\alpha}:=\int_{U_{\alpha,\text{red}}}\prod_i\dd x^i f_{\text{full}}(x)
\end{align}
where we expand  $\left.\Omega\right|_{U_\alpha}$ as
\begin{align}
	\left.\Omega\right|_{U_\alpha}=\left(f(x)+f_\mu(x)\theta^\mu%+\sum_{\mu<\nu}f_{\mu\nu}(x)\theta^\mu\theta^\nu
	+\cdots +f_{\text{full}}(x)\prod_\mu\theta^\mu\right)\prod_i\delta(\dd x^i)\prod_\mu\delta(\dd \theta^\mu).
\end{align}
With this definition, we can check the validity of various natural properties that integration operations should have, for instance the super-analog of Stokes' theorem.

For $\Omega$ to be globally defined, $\left.\Omega\right|_{U_\alpha}$ and $\left.\Omega\right|_{U_\beta}$ should be glued together by gluing function gluing $U_\alpha$ and $U_\beta$.
Note that this consistency does not always mean the gluing consistency of the locally defined top form $f_\text{full}\prod_i\dd x^i$ on the ordinary manifold $M_\text{red}$. A concrete example can be found in subsection \ref{effect}.

$\mathcal{N}=1$ super Riemann surfaces  are complex supermanifolds of dimension $1|1$ with odd deformation parameters $\eta$ with gluing functions of a particular form given by 
\begin{equation}
\begin{aligned}
	z^\prime&=u(z|\eta)+\theta\zeta(z|\eta)\sqrt{u(z|\eta)},\\
	\theta^\prime&=\zeta(z|\eta)+\theta\sqrt{\partial_zu(z|\eta)+\zeta(z|\eta)\partial_z\zeta(z|\eta)}.
\end{aligned}
	\label{sconf}
\end{equation}
This is the constraint so that the superconformal transformation generates the coordinate change as defined in  (\ref{sconf}).

The supermoduli space $\mathcal{M}$ is the complex supermanifold which parameterizes isomorphism classes of super Riemann surfaces of certain genus. The dimension of $\mathcal{M}$ is $3g-3|2g-2$ for $g\ge2$.

Super Riemann surfaces which can be constructed without any $\zeta$'s are called split.
For a split super Riemann surface, the transformation (\ref{sconf}) means that $\theta$ transforms as a section of a spin bundle $T\Sigma_\text{red}^{1/2}$ of the reduced Riemann surface $\Sigma_\text{red}$. So, the reduced space $\mathcal{M}_\text{red}$ of the supermoduli $\mathcal{M}$ is isomorphic to the moduli space  $\mathcal{M}_\text{spin}$ of ordinary Riemann surfaces with spin structure.

Super Riemann surfaces with punctures represents superstring worldsheets with vertex operators.
There are two types of punctures, NS punctures and R punctures. The dimension of the moduli space $\mathcal{M}_{n_\text{NS},n_\text{R}}$ of Super Riemann surfaces with $n_\text{NS}$ NS punctures and $n_\text{R}$ punctures is $3g-3+n_\text{NS}+n_\text{R}|2g-2+n_\text{NS}+n_\text{R}/2$.

To compute superstring amplitudes, 
we first define the moduli space for the right mover $\mathcal{M}_R$ and that for the left mover $\mathcal{M}_L$.
For type II theory, both are supermoduli spaces and for heterotic theory one is a supermoduli space and the other is an ordinary moduli space. 
Then we define a integration cycle $\Gamma \subset \mathcal{M}_L\times\mathcal{M}_R$ such that $\Gamma_\text{red}$ equals the diagonal $\cM_\red$ of $\mathcal{M}_{R,\text{red}}\times \mathcal{M}_{L,\text{red}}$. 
Then superstring amplitudes are given by an integral of a top form on $\Gamma$.
The choice of $\Gamma$ is not canonical, but integration on $\Gamma$ is.

\subsection{Superstring amplitudes as  integrals over the supermoduli}

Superstring perturbation theory is given by an integral over the super moduli space of a top form 
\begin{align}
	F_{\mathcal{V}}(\mathcal{J},\delta\mathcal{J}):=\int\mathcal{D}(\text{matter, ghost})\exp(-\widehat{I})\mathcal{V} 
	\label{ef}
\end{align} involving 
the action $\widehat I$ and the product of vertex operators $\mathcal{V}=\prod_i\mathcal{V}_i$ 
where  each vertex operator $\mathcal{V}_i$ is an unintegrated vertex operator of conformal dimension 0 and picture number $-1$ for an NS vertex operator and $-1/2$ for an R vertex operator.
Here the action $\widehat I$ is obtained as 
\begin{align}
	\widehat{I}=I+\frac{1}{2\pi}\int_{\Sigma}\mathcal{D}(\tilde{z},z|\tilde{\theta},\theta)\left(\delta\mathcal{J}B-\delta\widetilde{\mathcal{J}}\widetilde{B}\right)
\end{align}
where $I$ is the original worldsheet action, $\delta\mathcal{J}$ and $\delta\widetilde{\mathcal{J}}$ are superfields representing differentials of supercomplex structure of the worldsheet, and $B$ and $\widetilde{B}$ are superfields of the holomorphic and antiholomorphic superghosts.

Let $\mathcal{V}$ consist of $n_{\text{R}}$ R vertex operators and $n_{\text{NS}}$ NS vertex operators in terms of right movers,
and $\widetilde{n}_{\text{R}}$ R vertex operators and $\widetilde{n}_\text{NS}$ NS vertex operators in terms of lift movers. 
Then $F_\mathcal{V}$  is a form on $\mathcal{M}_{R,n_{\text{R}},n_{\text{NS}}}\times \mathcal{M}_{L,\widetilde{n}_\text{R},\widetilde{n}_\text{NS}}$.
The amplitude of the vertices $\mathcal{V}$ is then 
\begin{align}
	\mathcal{A}_\mathcal{V} := \frac{1}{2^g}\int_{\Gamma}^{}F_{\mathcal{V}}.
	\label{ampli}
\end{align}
The factor $\frac{1}{2^g}$ comes from GSO projection.

For heterotic string theory, the result is similar except for that the supermoduli space $\mathcal{M}_{L}$ is replaced by the bosonic moduli space.

\subsection{Coordinates of the supermoduli space}
One way to calculate the amplitude (\ref{ampli}) is to take an explicit coordinate on supermoduli space $\mathcal{M}_{g,n_{\text{NS}},n_{\text{R}}}$  with dimension $\Delta_e|\Delta_o$ where \begin{equation}
\Delta_e=3g-3+n_\text{NS}+n_\text{R},\qquad 
\Delta_o = 2g-2+n_{\text{NS}}+n_{\text{R}}/2.
\end{equation}
To study odd coordinates, consider a split super Riemann surface $\Sigma$.
Odd deformations from $\Sigma$ can be identified with $\chi \in H^1(\Sigma_{\text{red}},\widehat{\mathcal{R}})$ 
where
\begin{equation}
\widehat{\mathcal{R}} = \mathcal{R}\otimes \mathcal{O}\left(-\sum_{i=1}^{n_{\text{NS}}}z_i\right),\qquad
\mathcal{R}^2\simeq T\Sigma_{\text{red}}\otimes \mathcal{O}\left(-\sum_{i=1}^{n_{\text{R}}}x_i\right).
\label{defR}
\end{equation}
We call the modes $\chi$  the gravitino backgrounds, as they come from the two-dimensional supergravity field which couples to superstring.
We then choose a particular basis $\{\chi^\sigma\}$ of $H^1(\Sigma_{\text{red}},\widehat{\mathcal{R}})$ and expand $\chi$ as
\begin{align}
	\chi=\sum_{\sigma=1}^{\Delta_o}\eta_{\sigma}\chi^{\sigma}.
\end{align}
The term in $\widehat I$ involving $\chi$ is now 
\begin{align}
	I_\eta = \sum_{\sigma=1}^{\Delta_o}\frac{\eta_{\sigma}}{2\pi}\int_{\Sigma_{\text{red}}}\dd ^2 z\chi^{\sigma}\widehat{S}
	\label{chiS}
\end{align}
where $\widehat S$ is the supercurrent twisted to live in $\Gamma(\widehat{\mathcal{R}}^{-1}\otimes K\Sigma_\red)$:
\begin{align}
	\widehat{S}=fS
	\label{twist}
\end{align}
where $f$ is a locally defined function which behaves $f(z)\simeq \sqrt{z-x_i}$ near an R vertex at $x_i$ and $f(z)\simeq z-z_i$ near an unintegrated NS vertex at $z_i$.  The choice of a branch of $f$ corresponds to the choice of a square root of \eqref{defR}.
In Type II theory, there is also a term involing both $\chi$ and $\widetilde{\chi}$.
The coupling between $\delta\mathcal{J}$ and $\beta$ is similarly given by 
\begin{align}
	I_{\dd \eta}=\sum_{\sigma=1}^{\Delta_o}\frac{\dd \eta_\sigma}{2\pi}\int_{\Sigma_{\text{red}}}\dd ^2 z\chi^{\sigma}\widehat\beta
\end{align}
where $\widehat\beta$ is defined as in \eqref{twist}.

Inserting these couplings into (\ref{ef}), we get
\begin{multline}
	F_\mathcal{V}(m;\dd \eta|\eta;\dd m)
	%&=\Braket{\prod_{i=1}^{n}\mathcal{V}_i\exp(-I_\eta)}_{X_m,(b_1,c_1)}\times \left(\text{superghost contribution}\right)\\
	=\left<\prod_{i=1}^n\mathcal{V}_i\exp\left(-\sum_{\sigma=1}^{\Delta_o}\frac{\eta_{\sigma}}{2\pi}\int_{\Sigma_{\text{red}}}\dd ^2 z\chi^{\sigma}\widehat S
	-\sum_{\sigma=1}^{\Delta_o}\frac{\dd \eta_{\sigma}}{2\pi}\int_{\Sigma_{\text{red}}}\dd ^2 z\chi^{\sigma}\widehat\beta\right)\right.\\
	\times \left.\exp\left(-\sum_{j=1}^{\Delta_e}\frac{\dd m_j}{2\pi}\int_{\Sigma_{\text{red}}}\dd ^2z\mu^j \widehat{b} \right)\times(\text{antiholomorphic part})\times (\chi\widetilde{\chi} \text{term}) \right>_{m,g}.
	\label{efnu2}
\end{multline}
Here $\Braket{}_{m,g}$ denotes the correlation function under the metric background corresponding to coordinates $m$ of the genus-$g$ bosonic moduli space $\mathcal{M}_{g,\text{red}}$ 
and $\{\mu_i\}$ is a basis of Beltrami differentials of $n_\text{NS}+n_\text{R}$ punctured Riemann surface.
$\widehat{b}=gb$ is twisted to live in $\Gamma(K\Sigma_\text{red}\otimes\mathcal{O}(\sum z_i)\otimes\mathcal{O}(\sum x_i))$ (i.e. $g\sim z-z_i$ near a NS or R vertex at $z_i$).

\subsection{Effect of the change of the coordinate system}\label{effect}
Consider two open subsets $U_1$ and $U_2$ of $\mathcal{M}$ related by 
 a superdiffeomorphism caused by a vector field $\theta y(\tilde{z},z)\partial_z$. 
 This changes the gravitino background as 
\begin{align}
	\chi \rightarrow \chi '=\chi+\tilde{\partial}y
	\label{transfchi}
\end{align}
This induces to the change of basis $\chi^\sigma$ as
\begin{align}
	\chi^\sigma \rightarrow \chi^{\sigma\prime}=\chi^\sigma+\tilde{\partial}y^\sigma
\end{align}
where  $\sum \eta_\sigma y^\sigma =y$.

The modes $\chi^\sigma$ and $\chi^{\sigma\prime}$ represent the same class of $H^1(\Sigma_\text{red},\widehat{\mathcal{R}})$ and the coordinate $\eta_\sigma$ does not change. But the vector field $\theta y(\tilde{z},z)\partial_z$ causes a change on the metric at second order:
\begin{align}
	h_{zz}&\rightarrow h_{zz}+y\chi 
	      =h_{zz}+\sum_{\sigma,\sigma '}\eta_\sigma \eta_{\sigma '}y^{\sigma^\prime} \chi^\sigma.
	      \label{metrictransf}
\end{align}
The metric $h$ determines the bosonic coordinates of the moduli, and therefore this superdiffeomorphism gives rise to a change of the coordinates, mixing odd and even moduli parameters.

This mixing of odd and even coordinates leads to the subtleties mentioned above.
Consider a dimension $1|2$ complex supermanifold $M$ and an integral \begin{equation}
	\int_M \omega
	\label{superint}
\end{equation} where
$\omega$ is locally defined in a patch $U$ as
\begin{align}
	\omega=(\gamma_0(t)+\gamma_2(t)\eta_1\eta_2)\dd t\delta(\dd \eta_1)\delta(\dd \eta_2).
\end{align}

If we integrate first on $\eta_1$ and $\eta_2$, it reduces to 
\begin{align}
\longrightarrow	\int_{U_\text{red}}\gamma_2(t)\dd t.
	\label{holint}
\end{align}

But if $M$ is not holomorphically projected, we need a coordinate  change of the form
\begin{equation}
	t'=t+a(t)\eta_1\eta_2,\quad
	\eta_1'=\eta_1,\quad
	\eta_2'=\eta_2.
\end{equation}
In the new coordinate system $(t'|\eta_1',\eta_2')$, the integrand $\omega$ is now 
\begin{align}
	\omega=(\gamma_0'(t')+\gamma_2'(t')\eta_1\eta_2)\dd t'\delta(\dd \eta_1')\delta(\dd \eta_2')	
	\label{omega}
\end{align}
with
\begin{align}
	\gamma_2'(t')=\gamma_2(t')-\partial_{t'}\left(a(t')\gamma_0(t')\right)
	\label{omegap}
\end{align}
and the integral \eqref{superint} reduces to \begin{equation}
\longrightarrow	\int_{U_\text{red}}\gamma'_2(t)\dd t.\label{holint2}
\end{equation}
Then the reductions \eqref{holint} and \eqref{holint2} differ by an integral of an exact term $\partial_{t}\left(a(t)\gamma_0(t)\right)\dd t$  on $U_\red$.
This causes difficulties when we try to combine local contributions from patches together. 
In that case we should go to projected coordinates on $M$ to define the integration precisely, and this procedure destroys the holomorphic factorization property of the integrand.

Complex supermanifolds are not projected in general, so holomorphically factorised integrands on a complex supermanifold do not reduce to holomorphically factorised integrands on its reduced manifold.

\section{Poincar\'e dual of the reduced manifold}\label{varpi}
The arguments in the previous section also points the way out. 
Namely, we have an unambiguous equality  \begin{equation}
\int_M \omega
=\int_{M_\red} f
\end{equation}
if
\begin{align}
	\omega = f\prod_{\mu=1}^{\text{odddim}M}\eta^\mu\delta(\dd \eta^\mu).
\end{align} 
More invariantly under the coordinate change, we state that the form locally defined as \begin{equation}
\varpi=\prod_i\eta_i\delta(\dd \eta_i)
\end{equation}  is well-defined globally, and is  the \emph{Poincar\'e dual} of $M_\text{red}\subset M$, since we have 
\begin{align}
	\int_M \alpha\wedge\varpi =\int_{M_\text{red}} \left.\alpha\right|_{M_\text{red}}
	\label{pdual}
\end{align}
for all differential forms $\alpha$ without $\delta(\dd \eta)$ so that the multiplication on the left hand side meaningful. 
$\left.\alpha\right|_{M_\text{red}}$ can be obtained ignoring all the terms containing $\eta$'s and $\dd \eta$'s from $\alpha$.

Let us check that the form $\varpi$ is invariant under the coordinate change.
Suppose that two patches are glued as $\eta^{\mu\prime}=\psi^\mu(x|\eta)$.
Then,
\begin{align}
	\prod_\mu\delta(\dd \eta^{\mu\prime})&=\prod_\mu\delta\left(\dd \psi^\mu(x|\eta)\right)\\
					    &=\prod_\mu\delta\left(\sum_i\partial_{i}\psi^\mu\dd x^i+\sum_\nu\partial_{\nu}\psi^\mu\dd \eta^\nu\right)\\
		       &=\exp\left(\sum_{i,\mu}\partial_i\psi^\mu\dd x^i\partial_\mu\right)\prod_\mu\delta\left(\sum_\nu\partial_\nu\psi^\mu\dd \eta^\nu\right)\\
	&=\exp\left(\sum_{i,\mu}\partial_i\psi^\mu\dd x^i\partial_\mu\right)\frac{1}{\text{Det}A}\prod_\mu(\dd \eta^\mu)
\end{align} where \begin{equation}
	A_{\mu\nu}:=\left.\partial_\mu\psi^\nu\right|_{\eta=0}.
\end{equation}
Similarly, we have 
\begin{equation}
	\prod_\mu\psi^\mu=\prod_\mu\left(\sum_\nu\partial_\nu\psi^\mu|_{\eta=0}\eta^\nu+\text{higher order in $\eta$'s}\right)
	=\text{Det}A\prod_\mu\eta^\mu.
\end{equation}
Therefore  we have 
\begin{align}
	\varpi^\prime&=\prod_\mu\eta^{\mu\prime}\delta(\dd \eta^{\mu\prime})\\
		       &=\prod_\mu\psi^\mu\exp\left(\sum_i\partial_i\psi^\mu\dd x^i\partial_\mu\right)\delta\left(\sum_\nu\partial_\nu\psi^\mu\dd \eta^\nu\right)\\
		       &=\text{Det}A\prod_\mu\eta^\mu\exp\left(\mathcal{O}(\eta)\right)\frac{1}{\text{Det}A}\prod_\mu\delta(\dd \eta^\mu)\\
		       &=\prod_\mu\eta^\mu\delta(\dd \eta^\mu)\\
		&=\varpi.
\end{align}

So, if we are to reduce superstring amplitudes to bosonic string amplitudes,
we should check that the form  ${F}_\mathcal{V}$ in \eqref{efnu2} to be integrated can be represented as $\omega_\text{red}\wedge\varpi$.
We call this phenomenon the saturation of $\eta$'s and $d\eta$'s. 
This saturation also guarantees that the shift of the bosonic part of the supercoordinate near the degenerate super Riemann surfaces by even nilpotent terms does not affect the computation. In particular, we can safely integrate the fermionic coordinates of NS vertices.

We will explain in Appendix~\ref{fail} that this condition, in the more traditional langauge of PCOs, implies that the correlation function is completely independent of the positions of the PCOs.

\section{Bosonic string amplitudes as superstring amplitudes}\label{BV}

In this section we study how the mechanism studied in the previous section manifests itself in the embedding $\mathcal{N}=0$ string theory to $\mathcal{N}=1$ superstring theory in \cite{Bershadsky:1993cx}.

The $\mathcal{N}=0$  theory has the matter part $X_m$ and the $(b,c)$ ghost system.
We then construct an $\mathcal{N}=1$ matter system consisting of the matter system $X_m$, 
 the shifted ghost system $(b_1,c_1)$ whose spin are $(3/2,-1/2)$, and the standard superghost system $(b,c)$ and $(\beta,\gamma)$.
The super Virasoro generators of the $\cN=1$ matter system are
\begin{align}
	S_{\text{mat}}&=b_1+c_1(T_m+\partial c_1b_1)+\frac{5}{2}\partial^2 c_1,\\
	T_{\text{mat}}&=T_m-\frac{3}{2}b_{1}\partial c_1-\frac{1}{2}\partial b_1  c_1+\frac{1}{2}\partial^2(c_1\partial c_1),
\end{align}  where $T_m$ is the stress energy tensor of $X_m$. 
The central charge of $T_\text{mat}$ is $15$.

The shifted ghost system $(b_1,c_1)$ and $(\beta,\gamma)$ system have the same spin and the opposite statistics. They  are expected to cancel and the whole system goes back to the original $\mathcal{N}=0$ system consisting of $X_m$ and the $(b,c)$ ghost.
We will see below that the integration of  $(b_1,c_1)$ and $(\beta,\gamma)$ gives $\varpi=\prod \eta\delta(\dd \eta)$ which is the Poincar\'e dual of the reduced moduli space in the supermoduli space. 

The correspondence between $\mathcal{N}=1$ vertex operators and $\mathcal{N}=0$ vertex operators 
is as follows.
Let $V_i$ be a dimension-1 vertex operator of $X_m$. Then $c_1V_i +\theta V_i$ is a dimension-$1/2$ superconformal primary of the $\mathcal{N}=1$ matter system $(X_m,(b_1,c_1))$. Then, we can construct an $\mathcal{N}=1$ BRST invariant NS operator $\mathcal{V}_i(\widetilde{z},z|\theta)=c\delta(\gamma)(c_1V_i+\theta V_i)$.
Let us compute a form ${F}_{\mathcal{V}}$ defined in (\ref{ef}). Denote $n$ is the number of vertex operators. 
Denote the product of the vertex operators by  $\mathcal{V}=\prod_{i=1}^{n}\mathcal{V}_i$ where each $\mathcal{V}_i$ is defined as above.
Here we use NS operators  $\mathcal{V}_i$ fixed at $(\widetilde{z}_i,z_i|\theta=0)$.
Then, $F_\mathcal{V}$ becomes
\begin{multline}
	F_\mathcal{V}(m;\dd \eta|\eta;\dd m)
	%&=\Braket{\prod_{i=1}^{n}\mathcal{V}_i\exp(-I_\eta)}_{X_m,(b_1,c_1)}\times \left(\text{superghost contribution}\right)\\
	=\\
	\left<\prod_{i=1}^n\mathcal{V}_i\exp\left(-\sum_{\sigma=1}^{2g-2+n}\frac{\eta_{\sigma}}{2\pi}\int_{\Sigma_{\text{red}}}\dd ^2 z\chi^{\sigma}\widehat{S}\right)\right.
	\exp\left(%-\sum_{\sigma=1}^{2g-2+n}\frac{\eta_{\sigma}}{2\pi}\int_{\Sigma_{\text{red}}}\dd ^2 z\chi^{\sigma}S_{\text{bc}}
	-\sum_{\sigma=1}^{2g-2+n}\frac{\dd \eta_{\sigma}}{2\pi}\int_{\Sigma_{\text{red}}}\dd ^2 z\chi^{\sigma}\widehat{\beta}\right)\\
	\times \left.\exp\left(-\sum_{i=1}^{3g-3+n}\frac{\dd m_i}{2\pi}\int_{\Sigma_{\text{red}}}\dd ^2z\mu^i \widehat{b} \right)\right>_{m,g}\times(\text{antiholomorphic part}).
	\label{efnu}
\end{multline}
Here, $\{\chi^\sigma\}$ is the basis of gravitino backgrounds as above.

Only terms in (\ref{efnu}) which have the $b_1c_1$ ghost number $2g-2$ are nonzero. 
The product of vertex operators 
$\prod\mathcal{V}_i=\prod c\delta(\gamma)c_1V_i$ has the $b_1c_1$ ghost number $-n$.
Therefore we need to provide the $b_1c_1$ ghost number $2g-2+n$ from 
the expansion of $\exp\left(-\sum_{\sigma=1}^{2g-2+n}\frac{\eta_{\sigma}}{2\pi}\int_{\Sigma_{\text{red}}}\dd ^2 z\chi^{\sigma}\widehat{S}\right)$.
The only possibility is 
\begin{equation}
	\prod_{\sigma=1}^{2g-2+n} \frac{-\eta_{\sigma}}{2\pi}\int_{\Sigma_{\text{red}}}\dd ^2 z\chi^{\sigma}\widehat{b}_1.
\end{equation}
From a similar consideration on the $bc$ ghost number we conclude that only the term 
\begin{equation}
	\prod_{i=1}^{3g-3+n} \frac{-\dd m_i}{2\pi}\int_{\Sigma_{\text{red}}}\dd ^2 z\mu^i\widehat{b}
\end{equation}
contributes in the expansion of 
\begin{equation}
	\exp\left(-\sum_{i=1}^{3g-3+n}\frac{\dd m_i}{2\pi}\int_{\Sigma_{\text{red}}}\dd ^2z\mu^i \widehat{b}\right).
\end{equation}

Next, we explicitly calculate the integration with $b_1$ and $\beta$ field and confirm that the result does not depend on choice of $\{\chi^\sigma\}$ and reproduce the bosonic string amplitude. Let us expand $b_1$ by modes:
\begin{align}
	\widehat{b}_1=\sum_{\alpha=1}^{2g-2+n}v_{\alpha}\widehat{b}_1^\alpha +\sum_\lambda w_\lambda \widehat{b}_1^\lambda.
\end{align}
where $\widehat{b}_1^\alpha$ are bosonic zero modes, $\widehat{b}_1^\lambda$ are bosonic non-zero modes and $v_\alpha$ and $w_\lambda$ are fermionic variables.
The field $b_1$  is allowed to have poles at $z_i$ where vertex operators $c_1$ sit. 
Therefore zero modes $\{b_1^\alpha\}$ include meromorphic ones that have poles at $z_i$. 
Equivalently, $\{\widehat{b}_1^\alpha\}$ is a basis of 
\begin{equation}
H^0(\Sigma_\text{red},T\Sigma_\text{red}^{-3/2}\otimes\mathcal{O}(\sum_i z_i))
\end{equation}
 which is dual to the space of gravitino background 
\begin{equation}
 H^1(\Sigma_\text{red},T\Sigma_{\text{red}}^{1/2}\otimes\mathcal{O}(-\sum_i z_i)).
\end{equation}
Note that the superghost $\beta$ is  allowed to have poles at $z_i$ because of presence of vertex operator $\delta({\gamma})$.
Then, we can expand $\widehat{\beta}$ by the same basis as $\widehat{b}_1$:
\begin{align}
	\widehat{\beta}=\sum_{\alpha=1}^{2g-2+n}\nu_{\alpha}\widehat{b}_1^\alpha +\sum_\lambda \omega_\lambda \widehat{b}_1^\lambda.
\end{align}
where $\nu_\alpha$ and $\omega_\lambda$ are bosonic variables.

Integration with zero mode factors $v_\alpha$ and $\nu_\alpha$ in (\ref{efnu}) gives the factor
\begin{multline}
	\int\dd ^{2g-2}v_\alpha\dd ^{2g-2}\nu_\alpha 
	\exp\left(-\sum_{\sigma,\alpha}\frac{\eta_\sigma}{2\pi}\int_{\Sigma_{\text{red}}}\dd ^2z\chi^\sigma v_{\alpha}\widehat{b}_1^\alpha
	-\sum_{\sigma,\alpha}\frac{\dd \eta_\sigma}{2\pi}\int_{\Sigma_{\text{red}}}\dd ^2z\chi^\sigma \nu_{\alpha}\widehat{b}_1^\alpha\right) \\
	=\text{Det}M\prod_\sigma \delta(\eta_\sigma) \frac{1}{\text{Det}M}\prod_s \delta(\dd \eta_s)
	=\prod_\sigma \delta(\eta_\sigma)\delta(\dd \eta_\sigma).\label{BVfoo}
\end{multline}
where  $M^{\sigma\alpha}:=\frac{1}{2\pi}\int_{\Sigma_{\text{red}}}\chi^\sigma \widehat{b}_1^\alpha$.
Now it is obvious that spurious singularity which occurs when the matrix $M$ degenerates does not remain in the last form because of the cancellation with the contribution of $b_1$ integration.

Hence, $F_\mathcal{V}$ is proportional to $\varpi=\prod_{\sigma=1}^{2g-2}\left(\eta_\sigma\delta(\dd \eta_\sigma)\right)$ and according to argument in Sec.~\ref{varpi}, the amplitude becomes
\begin{align}
	\mathcal{A}_\mathcal{V}=\frac{1}{2^g}\int_{\Gamma_{n_\text{NS},n_\text{R},\text{red}}}\Braket{\prod_i\mathcal{V}_i\prod_j\int\dd ^2z\mu^j \widehat{b}}_\text{bosonic}.
	\label{Abos}
\end{align}

The final point to consider is that the integration space in (\ref{Abos}) is not the integration space of the bosonic string, because $\Gamma_{n_\text{NS},n_\text{R},\text{red}}=\mathcal{M}_{\text{spin}}$ is the moduli space of Riemann surfaces with spin structures. This point is resolved as follows \cite{Berkovits:1993xq}.
The moduli space of spin Riemann surfaces $\mathcal{M}_\text{spin}$ has two connected components  $\mathcal{M}^+$ and $\mathcal{M}^-$, the moduli space of Riemann surfaces with even and odd spin structure respectively. We can define a phase of amplitude to each connected component separately. A Riemann surface with genus $g$ has $2^{g-1}(2^g+1)$ even spin structures and $2^{g-1}(2^g-1)$ odd structure. 
We should give a factor of $-1$  to odd spin structure, then we have 
\begin{equation}
	\mathcal{A}_\mathcal{V}=\int_{\mathcal{M}_{\text{bosonic}}}\Braket{\prod_i\mathcal{V}_i\prod_j\int\dd ^2z\mu^j \widehat{b}}_\text{bosonic}.
\end{equation}

\section{Embedding topological string to superstring}\label{topological}
Let us now move on to the study of the graviphoton amplitudes in type II string theory
and its reduction to the topological string amplitude, which is a bosonic string theory as far as the integration over the moduli space is concerned. 

Let us consider the compactification of Type IIA/B theory with $\cN=2$ superconformal symmetry with central charge $c=9$. For definiteness we take Type IIA theory. 
We  Wick-rotate the time direction and introduce complex coordinates \begin{equation}
X_u=X_1+iX_2,\quad X_v=X_3+iX_4.
\end{equation}
 We use $\cN=(1,1)$ superfield $\mathscr{X}^\mu=X^\mu+\theta\psi^\mu+\widetilde{\theta}\widetilde{\psi}^\mu+\theta\widetilde{\theta}F^\mu$.
Here and in the following $\mu$ runs over $1,2,3,4$.
With this set up, the worldsheet supercurrents are given by
\begin{equation}
	S=i\psi_\mu\partial X^\mu + G^+ + G^-,\qquad
	\widetilde{S}=i\widetilde{\psi}_\mu\widetilde{\partial} X^\mu + \widetilde{G}^+ + \widetilde{G}^-.
 \end{equation}

\def\pp{\mathsf{p}}
\def\qq{\mathsf{q}}

We would like to consider amplitudes among 
$g-1$ graviphotons of momentum $p^\mu$, 
$g-1$ graviphotons of momentum $q^\mu$, 
one graviton of momentum $\pp^\mu$, 
and one graviton of momentum $\qq^\mu$.
We choose the polarizations so that graviton vertices $V_{R}$ and the graviphoton vertices $V_T$ are
\begin{equation}
\begin{aligned}
	V_{R,uvuv}&=
	c\widetilde{c}\delta(\gamma)\delta(\widetilde{\gamma})
	D_\theta\mathscr{X}^uD_{\widetilde{\theta}}\mathscr{X}^ue^{i(\pp_u\mathscr{X}^u+\pp_v\mathscr{X}^v)},\\
	V_{R,\bar{u}\bar{v}\bar{u}\bar{v}}&=
	c\widetilde{c}\delta(\gamma)\delta(\widetilde{\gamma})
	D_\theta\mathscr{X}^{\bar{v}}D_{\widetilde{\theta}}\mathscr{X}^{\bar{v}}e^{i(\qq_{\bar{u}}\mathscr{X}^{\bar{u}}+\qq_{\bar{v}}\mathscr{X}^{\bar{v}})},\\
	V_{T,uv}&=p_vc\widetilde{c}\Theta\widetilde{\Theta} \cS_1\widetilde{\cS}_1e^{ip_uX^u+ip_vX^v}\Sigma\widetilde\Sigma,\\
	V_{T,\bar{u}\bar{v}}&=q_{\bar{u}}c\widetilde{c}\Theta\widetilde{\Theta} \cS_2\widetilde{\cS}_2e^{iq_{\bar{u}}X^{\bar{u}}+iq_{\bar{v}}X^{\bar{v}}}\Sigma\widetilde\Sigma.
\end{aligned}
\end{equation}
Here $\cS_1$ and $\cS_2$ are spin fields which have charges $(1/2,1/2)$ and $(-1/2,-1/2)$ under the bosonized currents of $\psi^u$ and $\psi^v$,
$\Theta$ is the spin field for $\beta\gamma$ system, and 
$\Sigma$ is the left-moving and the right-moving vertex operators of the internal system which has $U(1)_R$ charge $(3/2,\mp 3/2)$ for typeIIA model and $(3/2,3/2)$ for typeIIB model constructed from the bosonized version of the $U(1)_R$ current. Here we suppose that metric and gravitino backgrounds are turned off near vertex operators.

We would like to check that there are no subtleties due to the odd moduli integration in the proof that the scattering amplitude of two gravitons and $2g-2$ graviphotons is equal to 
 the genus-$g$  topological vacuum amplitude $\mathcal{F}^\text{top}_{g}$:  \cite{Antoniadis:1993ze,Bershadsky:1993cx}
\begin{align}
	\mathcal{A}_{g}\left(V_{R,uvuv}V_{R,\bar{u}\bar{v}\bar{u}\bar{v}}(V_{T,uv})^{g-1}(V_{T,\bar{u}\bar{v}})^{g-1}\right)
	=V_{\mathbb{R}^4}\pp_v^2p_v^{g-1}\qq_{\bar{u}}^2q_{\bar{u}}^{g-1} \, (g!)^2\mathcal{F}^\text{top}_{g}
	\label{topamp}
\end{align}  to the leading order in the zero momentum limit.
We will see the mechanism of Sec.~\ref{varpi} at work again.

Let us denote the amplitude in the right hand side of (\ref{topamp}) as $\mathcal{A}$.
It has $2g-2$ RR vertices and 2 NS vertices, so the complex dimension of the supermoduli space is $5g-3|3g-1$.
Note that here we use unintegrated NS vertices.
We write $\mathcal{A}$ as 
\begin{align}
	\mathcal{A}=\frac{1}{2^g}\int_{\Gamma}F.
\end{align}
The integrand $F$ is
\begin{align}
	\nonumber F=&\left<\prod_iV_{T,uv}(x_i,\widetilde{x}_i)\prod_jV_{T,\bar{u}\bar{v}}(y_j,\widetilde{y}_i)V_{R,uvuv}(z,\widetilde{z}|\theta_1,\widetilde{\theta}_1)V_{R,\bar{u}\bar{v}\bar{u}\bar{v}}(w,\widetilde{w}|\theta_2,\widetilde{\theta}_2)\right.\\
		    &\times\left.\exp\left(-\sum_{\sigma=1}^{3g-1}\frac{\eta_{\sigma}}{2\pi}\int_{\Sigma_{\text{red}}}\dd ^2 z\chi^{\sigma}\widehat{S}-\sum_{\sigma=1}^{3g-1}\frac{\dd \eta_{\sigma}}{2\pi}\int_{\Sigma_{\text{red}}}\dd ^2 z\chi^{\sigma}\widehat{\beta}-\sum_{i=1}^{5g-3}\frac{\dd m_i}{2\pi}\int_{\Sigma_{\text{red}}}\dd ^2z\mu^i \widehat{b} \right)\right.\nonumber \\
	    \nonumber&\times\left.\exp\left(-\sum_{\sigma=1}^{3g-1}\frac{\widetilde{\eta}_{\sigma}}{2\pi}\int_{\Sigma_{\text{red}}}\dd ^2 z\widetilde{\chi}^{\sigma}\widehat{\widetilde{S}}-\sum_{\sigma=1}^{3g-1}\frac{\dd \widetilde{\eta}_{\sigma}}{2\pi}\int_{\Sigma_{\text{red}}}\dd ^2 z\widetilde{\chi}^{\sigma}\widehat{\widetilde{\beta}}-\sum_{i=1}^{5g-3}\frac{\dd \widetilde{m}_i}{2\pi}\int_{\Sigma_{\text{red}}}\dd ^2z\widetilde{\mu}^i \widehat{\widetilde{b}} \right)\right.\\
				   & \times\left.\exp \left( -\sum_{\sigma,\sigma'}\frac{\eta_\sigma\widetilde{\eta}_{\sigma'}}{4\pi} \int_{\Sigma_\text{red}}\dd ^2z (\widetilde{\psi}^\mu {\psi}_\mu + A^{++}+A^{+-}+A^{-+}+A^{--} ) \chi^\sigma \widetilde{\chi}^{\sigma'} \right)\right>_{m,g}.
	\label{efu}
\end{align}

$\chi^\sigma\in H^1(\sigma_{\text{red}},\widehat{\mathcal{R}})$ are the gravitino basis. $\{\mu_i\}$ is a basis of Beltrami differentials of Riemann surfaces with $2g$ marked punctures. 
$A^{\pm\pm}$ is an operator of internal theory which couples to $\chi^\mp\widetilde{\chi}^\mp$ to complete $\mathcal{N}=(2,2)$ local supersymmetry. Note that $A^{\pm\pm}$ have $\pm 1$ charge of holomorphic and antiholomorphic internal $U(1)_R$ symmetry because they couple to internal gravitino.

Let us first discuss the internal $U(1)_R$ charge.
The field $V_T$ has $(3/2, \mp3/2)$ internal $U(1)_R$ charge. So we need to bring down sufficient number of operators from second, third and fourth line in (\ref{efu}) to saturate $U(1)_R$ charge to get a nonzero contribution. All operators in second to third line in (\ref{efu}) have 0 or $\pm1$ $U(1)_R$ charge. So we should bring down $3g-3$ $\eta$'s and $3g-3$ $\widetilde{\eta}$.

Second, consider the terms involving $V_R$. Recall again that we use unintegrated vertices here; the degrees of freedom of the bosonic and fermionic positions are counted in the Beltrami differentials $\mu$'s and $\chi$'s. We therefore pick a specific value of the supercoordinates of the NS vertices, which we take to be $\theta_1=\theta_2=0,\widetilde{\theta}_1=\widetilde{\theta}_2=0$ for simplicity. Then, vertex operators become 
\begin{align}
	V_{R,uvuv}(\theta_1=0)=c\widetilde{c}\delta(\gamma)\delta(\widetilde{\gamma})\psi^u\widetilde{\psi}^ue^{i(\pp_uX^u+\pp_vX^v)},\\
	V_{R,\bar{u}\bar{v}\bar{u}\bar{v}}(\theta_2=0)=c\widetilde{c}\delta(\gamma)\delta(\widetilde{\gamma})\psi^{\bar{v}}\widetilde{\psi}^{\bar{v}}e^{i(\qq_{\bar{u}}X^{\bar{u}}+\qq_{\bar{v}}X^{\bar{v}})}.
\end{align}
Operators which has the charge of $\psi^u$ are the following: $\psi^u$ itself in $V_R$, $\cS_{1,2}$ in $V_T$ and $\psi^u,\psi^{\bar{u}}$ in $\chi S$ term. Therefore, to have a nonzero contribution, we need to use the $\chi S$ term.
We can treat   $\psi^{\bar{v}},\widetilde{\psi}^u$ and $\widetilde{\psi}^{\bar{v}}$ in a similar manner. 
In total, we need two $\eta$'s and two $\widetilde{\eta}$'s to saturate these charges.

Considering both the charge of $U(1)_R$ and the charges of $\psi$'s, we see that the only term  which contributes has $3g-1$ of $\eta$'s and $3g-1$ of $\widetilde{\eta}$'s.   
Therefore, it involves all the odd moduli, and the  amplitudes we are considering can be represented as an integral over the ordinary bosonic moduli space, by the mechanism of Sec.~\ref{varpi}. 
We can safely integrate the odd moduli.

Two of $\eta$'s and two of $\tilde\eta$'s  correspond to the fermionic positions of unintegrated NSNS vertices.
We integrate these odd moduli first to convert them to the integrated NSNS vertices.
Then we have 
\begin{align}
	\mathcal{A}=\frac{1}{2^g}\int_{\Gamma'}F_\text{int}
\end{align}
where $\Gamma'$ is a integration cycle in the supermoduli space of super Riemann surfaces with $2g-2$ RR punctures whose dimension is $5g-5|3g-3$.  The integrand $F_\text{int}$ is
\begin{align}
	\nonumber F_\text{int}=&\left<\prod_iV_{T,zw}(x_i)\prod_jV_{T,\bar{u}\bar{v}}(y_j)\int_{\Sigma_{\text{red}}}\dd ^2zV^\prime_{R,uvuv}\int_{\Sigma_\text{red}}\dd w V^\prime_{R,\bar{u}\bar{v}\bar{u}\bar{v}}\right.\\
			   &\times\left.\prod_{\sigma=1}^{3g-3}\frac{\eta_{\sigma}}{2\pi}\int_{\Sigma_{\text{red}}}\dd ^2 z\chi^{\sigma}\widehat{G}^-\prod_{i=1}^{5g-5}\frac{\dd m_i}{2\pi}\int_{\Sigma_{\text{red}}}\dd ^2z\mu^i \widehat{b}\;\; \exp\left(-\sum_{\sigma=1}^{3g-3}\frac{\dd \eta_{\sigma}}{2\pi}\int_{\Sigma_{\text{red}}}\dd ^2 z\chi^{\sigma}\widehat{\beta}\right)\right.\nonumber \\
	     &\times\left.\prod_{\sigma=1}^{3g-3}\frac{\widetilde{\eta}_{\sigma}}{2\pi}\int_{\Sigma_{\text{red}}}\dd ^2 z\widetilde{\chi}^{\sigma}\widehat{\widetilde{G}}^\pm\prod_{i=1}^{5g-5}\frac{\dd \widetilde{m}_i}{2\pi}\int_{\Sigma_{\text{red}}}\dd ^2z\widetilde{\mu}^i \widehat{\widetilde{b}}\;\;\exp\left(-\sum_{\sigma=1}^{3g-3}\frac{\dd \widetilde{\eta}_{\sigma}}{2\pi}\int_{\Sigma_{\text{red}}}\dd ^2 z\widetilde{\chi}^{\sigma}\widehat{\widetilde{\beta}}\right) \right.\nonumber\\
	     &+ \left.(\text{terms involving $A\chi\widetilde{\chi}$})\;\; \right>_{m,g}
	\label{efu2}
\end{align}
where $V_R^\prime$ is the picture number 0 NSNS vertex:
\begin{equation}
\begin{aligned}
	V_{R,uvuv}^\prime&=\left(\partial X^u\widetilde{\partial} X^u-\pp_v^2\psi^u\psi^v\widetilde{\psi}^u\widetilde{\psi}^v\right)e^{i(\pp_uX^u+\pp_vX^v)},\\
	V_{R,\bar{u}\bar{v}\bar{u}\bar{v}}^\prime&=\left(\partial X^{\bar{v}}\widetilde{\partial} X^{\bar{v}}-\qq_{\bar{u}}^2\psi^{\bar{u}}\psi^{\bar{v}}\widetilde{\psi}^{\bar{u}}\widetilde{\psi}^{\bar{v}}\right)e^{i(\qq_{\bar{u}}X^{\bar{u}}+\qq_{\bar{v}}X^{\bar{v}})}.
\end{aligned}
\end{equation}
Now $m,\eta,\mu,\chi$ and twisted fields $\widehat{b},\widehat{G},\widehat{\beta}$ and their antiholomorphic counterparts are appropriate ones for $\Gamma'$.

Then we integrate the $\eta$ directions of $\Gamma'$, resulting in 
\begin{align}
	\mathcal{A}=&\frac{1}{2^g}\int_{\Gamma_{\text{red}}}F_\text{red}
\end{align}
where
\begin{align}
	\nonumber F_\text{red}=&\left<\prod_iV_{T,uv}(x_i)\prod_jV_{T,\bar{u}\bar{v}}(y_j)\int_{\Sigma_{\text{red}}}\dd ^2zV^\prime_{R,uvuv}\int_{\Sigma_\text{red}}\dd ^2w V^\prime_{R,\bar{u}\bar{v}\bar{u}\bar{v}}\right.\\
	&\times\left.\prod_{\sigma=1}^{3g-3}\int_{\Sigma_{\text{red}}}\dd ^2 z\chi^{\sigma}\widehat{G}^-\prod_{\sigma=1}^{3g-3}\delta\left(\int_{\Sigma_{\text{red}}}\dd ^2 z\chi^{\sigma}\widehat{\beta}\right)\prod_{i=1}^{5g-5}\frac{\dd m_i}{2\pi}\int_{\Sigma_{\text{red}}}\dd ^2z\mu^i \widehat{b}\;\; \right.\nonumber \\
	&\times\left.\prod_{\sigma=1}^{3g-3}\int_{\Sigma_{\text{red}}}\dd ^2 z\widetilde{\chi}^{\sigma}\widehat{\widetilde{G}}^\pm\prod_{\sigma=1}^{3g-3}\delta\left(\int_{\Sigma_{\text{red}}}\dd ^2 z\widetilde{\chi}^{\sigma}\widehat{\widetilde{\beta}}\right) \prod_{i=1}^{5g-5}\frac{\dd \widetilde{m}_i}{2\pi}\int_{\Sigma_{\text{red}}}\dd ^2z\widetilde{\mu}^i \widehat{\widetilde{b}}\right.\nonumber\\
	     &+ \left.(\text{terms involving $A\chi\widetilde{\chi}$})\;\; \right>_{m,g}
\end{align}

At this stage we can set $\chi^\sigma$ to have delta function support at appropriate places, at least on generic points of $\Gamma_\text{red}$. Then we go to the familiar picture changing formalism.
As reviewed in Appendix.~\ref{fail}, 
the saturation as above guarantees that the correlation function does not depend at all on the positions of the PCOs.\footnote{When the internal CFT is a free CFT, such as orbifolds of $T^6$, this independence from the positions of the PCOs implies a host of  intricate identities among higher-genus theta functions. The authors have not been able to prove these identities by themselves. Rather, they regard these identities as dervied via the CFT methods. } Therefore we can place the PCOs at the points most suitable for calculations. 
First, the terms involing $A$ vanish if we place holomorphic and antiholomorphic PCOs at distinct points.
Then, we choose the place the PCOS as in the calculation of \cite{Antoniadis:1993ze}. 
The rest of the computation goes unchanged compared to \cite{Antoniadis:1993ze}. 
To briefly summarize, we  explicitly evaluate the contributions from the spacetime bosons and fermions, the ghosts $(b,c)$ and $(\beta,\gamma)$, and the internal $U(1)_R$ boson.
Then the integral over $x_i$, $y_i$ and Beltrami differentials associated to them can also be performed.
We end up with\begin{equation}
F_\text{red}=\pp_v^2 p_v^{g-1} \qq_{\bar u}^2 q_{\bar u}^{g-1} V_{\mathbb{R}^4}(g!)^2 \left<\prod_{i=1}^{3g-3}\frac{\dd m_i}{2\pi}\int_{\Sigma_{\text{red}}}\dd ^2z\mu^i G^-\prod_{i=1}^{3g-3}\frac{\dd \widetilde{m}_i}{2\pi}\int_{\Sigma_{\text{red}}}\dd ^2z\widetilde{\mu}^i \widetilde{G}^-\right>_{m,g}
\end{equation}  at the leading order of momenta, 
where $G$, $\widetilde{G}$ are now topologically twisted.
This is the relation \eqref{topamp} we wanted to show.

\paragraph{Acknowledgments}
The authors thank Edward Witten for helpful discussions. 
They also thank their colleagues who spent tens of hours attending the journal club  by KO on  the contents of \cite{Witten:2012bg,Witten:2012ga,Witten:2012bh}.
YT thanks the hospitality of the Institute for Advanced Study where this manuscript was finalized.
The work of KO is partially supported by an Advanced Leading Graduate Course for Photon Science grant.
The work of YT is partially supported by World Premier International Research Center Initiative
(WPI Initiative),  MEXT, Japan through the Institute for the Physics and Mathematics
of the Universe, the University of Tokyo.

\eject

\appendix

\section{Formalism using PCOs}\label{fail}
In this appendix, we review the relationship between the approach using the supermoduli space and the approach which uses the picture changing operators (PCOs). 
\subsection{Pointlike gravitinos and PCOs}
We start from the integrand $F(m,\dd \eta|\eta,\dd m)$ \eqref{efnu2} of a general superstring amplitude. 
Let us explicitly perform the integration over $\eta$ directions, which produces factors of the form
\begin{multline}
	\prod_{\sigma}^{}
	\delta\left(\int_{\Sigma_{\text{red}}}\dd ^2z \chi^\sigma \widehat{\beta}\right)
	\delta\left(\int_{\Sigma_{\text{red}}}\dd ^2z \chi^\sigma \widehat{S}\right)
	\delta\left(\int_{\Sigma_{\text{red}}}\dd ^2z \widetilde{\chi}^\sigma \widehat{\widetilde{\beta}}\right)
	\delta\left(\int_{\Sigma_{\text{red}}}\dd ^2z \widetilde{\chi}^\sigma \widehat{\widetilde{S}}\right)\\
	+(\text{contribution from $\chi\widetilde{\chi}$ term}).
	\label{chifactor}
\end{multline}
A traditional choice of the gravitino basis $\{\chi\}$ is to take $\chi^\sigma = \delta(p_\sigma),\widetilde{\chi}^\sigma = \delta(q_\sigma)$ for some $p_\sigma\,q_\sigma in\Sigma_{\text{red}}$.
If all $p_\sigma$ and $q_\sigma$ are different, $\chi\widetilde{\chi}$ term does not contribute.

Then, the holomorphic part of the factors (\ref{chifactor}) above becomes
\begin{align}
	\prod_{\sigma}\mathscr{Y}(p_\sigma)=
	\prod_{\sigma}
	\delta\left(\widehat{\beta}(p_\sigma)\right)
	\widehat{S}(p_\sigma)=
	\prod_{\sigma}
	\delta\left(\beta(p_\sigma)\right)
	S(p_\sigma)\label{Ydef}
\end{align}
 where the operator $\mathscr{Y}(p_\sigma):=	\delta\left(\widehat{\beta}(p_\sigma)\right)\widehat{S}(p_\sigma)$ is  the PCO.
Summarizing, we have
\begin{multline}
	 \int_U  F(m,\dd\eta|\eta,\dd m) =\\
	  \int_{U_\text{red}}\Braket{\prod_i\mathcal{V}_i\prod_{\sigma=1}^{\Delta_o}\mathscr{Y}(p_\sigma)\prod_{j=1}^{\Delta_e}\left(\dd m^j\int_{\Sigma_\text{red}}\dd ^2z\mu^j\widehat{b}\right)}\times (\text{antiholomorphic part})
	\label{picture}
\end{multline} 
where $U$ is a patch in the supermoduli space, and $U_\text{red}$ is the corresponding patch in the bosonic moduli space.
In the following we do not explicitly write down the antiholomorphic part, as it can be dealt with separately.

There are two points to be kept in mind when this formula is used.
The first point is that we can not always guarantee that $\chi^\sigma$ are linearly independent. 
Namely, we assumed above that  $[\chi^\sigma]=[\delta(p_{\sigma})]$ forms a basis of $H^1(\Sigma_\text{red},\widehat{\mathcal{R}})$,
where [{\small\textbullet}] denotes the class in $H^1(\Sigma_\text{red},\widehat{\mathcal{R}})$. 
This is not always the case.  To see this, let us write down the condition when $[\delta(p_\sigma)]$ are linearly  dependent in $H^1(\Sigma_\text{red},\widehat{\mathcal{R}})$, namely that the equation
\begin{align}
	\widetilde{\partial}y=\sum_{\sigma=1}^{\Delta_o} e_\sigma\delta(p_\sigma)
	\label{pccond}
\end{align}
has solutions for some complex numbers $e_\sigma$ and $y$.
This means that $y$ has meromorphic and has poles only at $p_\sigma$'s. Equivalently, 
\begin{equation}
\dim H^0(\Sigma_\text{red},\widehat{\mathcal{R}}\otimes \mathcal{O}(\sum_\sigma p_\sigma)) >0.\label{pcfail}
\end{equation}
When this happens the superghost correlation function has a pole, which is called the spurious singularity in the literature. Its appearance can be seen e.g.~in \eqref{BVfoo}.

\subsection{Dependence of the correlators on the positions of the PCOs}
Another point to be kept in mind is more severe: the change in the positions of the PCOs generate exact forms on the bosonic moduli space $M_\red$, as a manifestation of the phenomenon described in Sec.~\ref{effect}. This is because the change in the positions of the PCOs is a change in the coordinate system of the supermoduli space. 
The following argument is based on the one given in \cite{Atick:1987wt}, and it is only a slight extension thereof. 
The change can be derived by manipulating the derivative of the correlator \eqref{picture} with the PCOs with respect to the positions $p_\sigma$ directly on a purely bosonic Riemann surface.
Equivalently, the same change can be derived by studying its effect on the coordinates on the moduli space of the super Riemann surfaces. We use this second point of view below, and compute 
\begin{equation}
\Delta \int_U  F(m,\dd\eta|\eta,\dd m) = 
	  \int_{U_\text{red}}\Delta \Braket{\prod_i\mathcal{V}_i\prod\mathscr{Y}(p_\sigma)\prod\left(\dd m^j\int_{\Sigma_\text{red}}\dd ^2z\mu^j\widehat{b}\right)} \label{input}
\end{equation} under the change of the positions of the PCOs.

Let us first put PCOs at $\{p_\sigma\}$, and and consider the effect of  moving one PCO at $p_1$ to $p_1+\Delta p_1$ and consider only first order in $\Delta p_1$.
We should find a superconformal gauge transformation parameter $y$ in \eqref{transfchi} producing this gauge transformation.
Explicitly, we should solve
\begin{align}
	\sum_{\sigma'}\eta'_{\sigma'}\delta(z-p'_{\sigma'})=\widetilde{\partial}y(z)+\sum_\sigma\eta_\sigma\delta(z-p_\sigma)
	\label{movePCO}
\end{align}
in $y$ and $\eta'$. Here $p'_\sigma = p_\sigma$ for $\sigma\neq1$ and $p'_1=p_1+\Delta p_1$.
This is a slight extension of the transformation dealt in Sec.~\ref{effect}, in that 
we allowed $\eta$'s to vary.
As we see below,  the transformation of $\eta$'s are linear, and thus the change in $\eta$s does not cause any effect, as it can be absorbed in a redefinition of $\eta$s.
%but this affect nothing because linear transformation of $\eta$'s can be absorbed by redefinition of $\eta$'s (we will see that this is a linear transformation).

To solve \eqref{movePCO}, we define a Green's function $G(z,w)$ with the property
\begin{align}
	\partial_{\widetilde{z}}G(z,w)=\delta(z-w)+\sum_\sigma R_\sigma(w)\delta(z-p'_\sigma).
	\label{green}
\end{align}  The terms $R_\sigma(w)$ are source terms necessary to solve the Laplace equation on a closed nontrivial Riemann surfaces.
Then, we can solve \eqref{movePCO} with
\begin{align}
	y(z)=-\int \dd ^2w G(z,w)\chi(w) = -\sum_\sigma G(z,p_\sigma)\eta_\sigma,
	\label{solutiony}\\
	\eta'_\sigma = -\int \dd ^2w R_\sigma(w)\chi(w) = -\sum_{\tau}R_\sigma(p_{\tau})\eta_{\tau}.
\end{align}

Considering that $y$ is a section of $\widehat{\mathcal{R}}$ and \eqref{movePCO},\eqref{solutiony}, we see that 
$G(z,w)$ should transforms as a section of $\widehat{\mathcal{R}}$ as a function of the $z$-plane
and as a section of $\widehat{\mathcal{R}}^{-1}\otimes K\Sigma_\text{red}$ as a function of the $w$-plane.

The Green's function \eqref{green}  can be written  as a correlator of twisted $\beta\gamma$ system:
\begin{align}
	G(z,w)&=\frac{1}{Z}\Braket{\widehat{\gamma}(z)\widehat{\beta}(w)\prod_\sigma \delta(\widehat{\beta}(p'_\sigma))}_{\widehat{\beta}\widehat{\gamma}}   ,
	\label{Gzw}\\
	Z&=\Braket{\prod_\sigma\delta(\widehat{\beta}(p'_\sigma))}_{\widehat{\beta}\widehat{\gamma}}.
	\label{Z}
\end{align}
Here, twitsted fields $\widehat{\beta}$ and $\widehat{\gamma}$ are defined as
\begin{align}
	\widehat{\beta}=f\beta, \quad \widehat{\gamma}=f^{-1}\gamma
\end{align}
using the same $f$ as in \eqref{twist}.  This means that $\widehat{\gamma}\in \Gamma(\widehat{\mathcal{R}})$ and $\widehat{\beta}\in\Gamma(\widehat{\mathcal{R}}^{-1}\otimes K\Sigma_\text{red})$.
Hence, the expression \eqref{Gzw} reproduces desired transformation laws and positions and residues of poles of $G(z,w)$.
More on twisted $\beta\gamma$ system can be found in \cite{Witten:2012bg}.

The superconformal transformation \eqref{movePCO} with $y(z)$ also causes the transformation of the metric as in \eqref{metrictransf}.
So, the bosonic moduli parameter $m$ transforms as
\begin{align}
	 \Delta m_i = \int \dd ^2z y(z)\chi(z) \widehat{b}_i = -\sum_{\sigma\neq 1}\eta_1\eta_\sigma G(p_\sigma,p_1) b_i(p_\sigma).
\end{align}
Here, $\{b_i\}$ is a basis of (twisted) quadratic differentials which is dual to a given basis $\{\mu^i\}$ of Beltrami differentials. We also used $G(z,p_\sigma)=0$.

Summarizing, the superconformal transformation with parameter $y(z)$ corresponds to a coordinate change of supermoduli space given by 
\begin{align}
	\eta_\sigma &\rightarrow \eta'_\sigma,&
	m_i &\rightarrow m'_i = m_i - \Delta m_i\label{kanter}
\end{align} where \begin{equation}
\Delta m_i=\sum_{\sigma\neq 1}\eta_1\eta_\sigma G(p_\sigma,p'_1) b_i(p_\sigma).
\end{equation}
Let us write the integrand of the superstring amplitude as \begin{equation}
\int_U F(m,\dd\eta|\eta,\dd m)=
\int_U \prod_i\dd m_i\prod_\sigma\delta(\dd \eta_\sigma) F(m|\eta).\label{shortF}
\end{equation}
Under the changes \eqref{kanter} this integral is changed according to 
\begin{align}
	\prod_i\dd m_i\prod_\sigma\delta(\dd \eta_\sigma) F(m|\eta) \rightarrow \prod_i\dd m'_i\prod_\sigma\delta(\dd \eta_\sigma) \left(F(m'|\eta)-\sum_i\partial_i \left(\Delta m F(m'|\eta)\right)\right)
	\label{shiftF}
\end{align}
to the first order of $\Delta p_1$. Therefore, the total change to the first order is \begin{align}
&\Delta\int_U F(m,\dd\eta|\eta,\dd m) \nonumber \\
&=
-\int_{U_\text{red}}
\prod_i\dd m'_i  \sum_i\partial_i \left[\Delta m_i F(m'|\eta)\right]_\text{full} \label{XXX} \\
&= \int_{U_\text{red}}
\prod_i\dd m'_i \sum_i \partial_i \left[\sum_\sigma (-1)^{\sigma-1} G(p_\sigma,p'_1)b_i(p_\sigma) (F(m|\eta)|_{1\sigma})\right],\label{YYY}\\
&= \int_{U_\text{red}}\partial_i \Bigl[ \sum_\sigma (-1)^{\sigma-1} b_i(p_\sigma) \Bigl< \widehat{\gamma}(p_\sigma)\widehat{\beta}(p'_1) \prod_{\tau}\delta(\widehat{\beta}(p_\tau))
\times\nonumber \\ & \hspace{3cm} 
	\prod_{\rho\neq 1,\sigma} \widehat{S}(p_\rho) 
	\prod_i\mathcal{V}_i\prod\left(\dd m^j\int_{\Sigma_\text{red}}\dd ^2z\mu^j\widehat{b}\right)\Bigr>\Bigr]\label{ZZZ} 
\end{align}
Here, in \eqref{XXX},   we denoted the coefficient of $\prod_{\tau} \eta_\tau$ in $A(m|\eta)$ as $A|_\text{full}$,
 and in \eqref{YYY},  the coefficient of $\prod_{\tau\neq 1,\sigma} \eta_\tau$ in $A(m|\eta)$  is denoted by as $A|_{1\sigma}$, and to go from \eqref{YYY} to \eqref{ZZZ}, we used the fact that 
there is a factor of $Z$ given in \eqref{Z} in the definition of $F$
which  cancels $Z$ in the denominator of the expression for $G(z,w)$ in \eqref{Gzw}.

Comparing with \eqref{input} and using the definition of the PCO \eqref{Ydef}, we finally find \begin{multline}
\Delta \Braket{\prod_i\mathcal{V}_i\prod\mathscr{Y}(p_\sigma)\prod\left(\dd m^j\int_{\Sigma_\text{red}}\dd ^2z\mu^j\widehat{b}\right)} 
= \\
\partial_i \Bigl[ \sum_\sigma (-1)^{\sigma-1} b_i(p_\sigma) \Bigl< \widehat{\gamma}(p_\sigma)\widehat{\beta}(p'_1) \delta(\widehat{\beta}(p_1))\delta(\widehat{\beta}(p_\sigma)) \times \\
	\prod_{\rho\neq 1,\sigma} \mathscr{Y}(p_\rho) 
	\prod_i\mathcal{V}_i\prod\left(\dd m^j\int_{\Sigma_\text{red}}\dd ^2z\mu^j\widehat{b}\right)\Bigr>\Bigr].\label{0000}\end{multline}
The right hand side is an  exact form on the patch of the  bosonic moduli space. 
This can also be obtained by a schematic manipulation as follows\cite{Friedan:1985ge,Verlinde:1987sd}:
To move a PCO $\mathscr{Y}(p)$ to another position $q$, we first go to the large Hilbert space and move the BRST operator: 
\begin{align}
\Braket{\mathscr{Y}(p)\cdots}
&=\Braket{[Q_\text{BRST}, \xi (p)] \xi(q) \cdots}\\
&=\Braket{\xi (p) [Q_\text{BRST}, \xi(q)] \cdots}
+\Braket{\xi (p) \xi(q)[Q_\text{BRST},  \cdots]}\\
&=\Braket{ \mathscr{Y}(q) \cdots}
+\Braket{\xi (p) \xi(q)[Q_\text{BRST},  \cdots]}.\label{PCO}
\end{align} The second term in the last line contains terms where $Q_\text{BRST}$ acts on the $b$ ghost, thus generating exact terms on the moduli space of Riemann surfaces. 

At this point, it is obvious that saturation of $\eta$'s guarantees that the correlator is independent of the positions of PCO's.
When $F(m,\dd\eta|\eta,\dd m)$ is proportional to $\varpi$, as discussed in Sec.~\ref{varpi}, 
$F(m|\eta)$ as defined in \eqref{shortF} only has terms with all factors of $\eta$, and
$F(m|\eta)|_{1\sigma}$ vanishes by definition.
Therefore,  the right hand side of \eqref{0000} also vanishes,  and the correlation function \eqref{picture} becomes completely independent of the positions of the PCOs.

Of course, we can check that independence directly by calculating \eqref{PCO}.
The derivation using the structure of the supermoduli space is conceptually useful to understand how and when the formalism using PCO's is convenient.

%Combining \eqref{shiftF}-\eqref{shiftF2} and considering $\gamma b$ is the supercurrent of the superghost system, 
%we get the same result as described in \eqref{PCO}.
%$\clubsuit$ mention this appendix around \eqref{PCO}! $\clubsuit$

%In the picture changing formalism, this phenomenon surfaces itself as the dependence of the correlators on the position of the PCOs, since changing their positions induces the coordinate changes of the super moduli space. Let us see it in a more traditional language. To move a PCO $\mathscr{Y}(p)$ to another position $q$, we first go to the large Hilbert space and move the BRST operator: 
%\begin{align}
%\Braket{\mathscr{Y}(p)\cdots}
%&=\Braket{[Q_\text{BRST}, \xi (p)] \xi(q) \cdots}\\
%&=\Braket{\xi (p) [Q_\text{BRST}, \xi(q)] \cdots}
%+\Braket{\xi (p) \xi(q)[Q_\text{BRST},  \cdots]}\\
%&=\Braket{ \mathscr{Y}(q) \cdots}
%+\Braket{\xi (p) \xi(q)[Q_\text{BRST},  \cdots]}.\label{PCO}
%\end{align} The second term in the last line contains terms where $Q_\text{BRST}$ acts on the $b$ ghost, thus generating exact terms on the moduli space of Riemann surfaces. 

%\newpage

\bibliographystyle{ytphys}
\bibliography{bosamp}
\end{document}